\documentclass[useAMS,usenatbib]{mn2e}
\usepackage{graphicx}
\usepackage{dcolumn}
\usepackage{bm}

\usepackage{amsmath}
\usepackage{amssymb}

\newcommand{\mm}{\mu_{max}}
\newcommand{\lo}{{l_\odot}}

\title[Extragalactic dark matter and direct detection]{Extragalactic dark matter and direct detection experiments}

\author[A. N. Baushev]{A. N. Baushev\thanks{E-mail:
baushev@gmail.com}\\
DESY, 15738 Zeuthen, Germany\\
 Institut f\"ur Physik und Astronomie, Universit\"at Potsdam, 14476
Potsdam-Golm, Germany\\}
\begin{document}

\date{}

\pagerange{\pageref{firstpage}--\pageref{lastpage}} \pubyear{2012}

\maketitle

\label{firstpage}

\begin{abstract}
Recent astronomical data strongly suggest that a significant part of the dark matter, composing the
Local Group and Virgo Supercluster, is not incorporated into the galaxy haloes and forms diffuse
components of these galaxy clusters. Apparently, a portion of the particles from these components
may penetrate into the Milky Way and make an extragalactic contribution to the total dark matter
containment of our Galaxy.

We find that the particles of the diffuse component of the Local Group are apt to contribute $\sim
12\%$ to the total dark matter density near the Earth. The particles of the extragalactic dark
matter stand out because of their high speed ($\sim 600$~{km/s}), i.e. they are much faster than
the galactic dark matter. In addition, their speed distribution is very narrow ($\sim 20$~{km/s}).
The particles have isotropic velocity distribution (perhaps, in contrast to the galactic dark
matter). The extragalactic dark matter should give a significant contribution to the direct
detection signal. If the detector is sensitive only to the fast particles ($v<450$~{km/s}), the
signal may even dominate.

The density of other possible types of the extragalactic dark matter (for instance, of the diffuse
component of the Virgo Supercluster) should be relatively small and comparable with the average
dark matter density of the Universe. However, these particles can generate anomaly high energy
collisions in direct dark matter detectors.
\end{abstract}


\section{Introduction}
It is widely believed that all the dark matter particles (hereafter DMPs), which a terrestrial
observer can detect, belong to the Milky Way Galaxy. The main aim of this letter is to dispute this
assertion and to show that a remarkable fraction of dark matter particles detected on the Earth
does not probably belong to our Galaxy. Although their density is relatively small, as compared
with the total dark matter density, their impact into the direct detection signal may even dominate
because of high speeds of the particles.

According to the modern cosmological notion, the haloes of giant galaxies, like Milky Way or
Andromeda, are regions of local dark matter overdensity, rather than isolated islands. Indeed, the
Local Group, along with the haloes of large and dwarf galaxies, contains a significant fraction of
dark matter that is not bound in the galaxies and presumably forms a large envelope of the Local
Group \citep{bt}. A significant part of the dark matter of the Virgo Supercluster is also not
localized in haloes and probably distributed more or less homogeneously over all the volume of the
Supercluster \citep{karach11}. Some part of this diffuse dark matter (preeminently from the Local
Group envelope) penetrates into the central region of our Galaxy and gives a contribution to the
direct detection signal, which can even dominate under certain conditions: as we will see, the
density fraction of the extragalactic dark matter is relatively small ($\sim 12\%$), however its
particles should have extremely high speeds, close to the escape velocity ($\sim 600$~{km/s}) or
even higher. It sets off the extragalactic particles from the halo DMPs with much lower average
speed. The direct detection signal produced by this component should also have some other
characteristic features that will be discussed below.

\section{The dark matter envelope of the Local Group}
Unfortunately, the total mass, distribution and dynamical properties of the extragalactic dark
matter environment are now poorly known. Therefore, we have to do with rough estimates of its
content near the Solar System.

The Local Group consists of two very massive galaxies (Milky Way and Andromeda galaxy M31), less
massive Triangulum galaxy M33, and a host of dwarf galaxies. It seems reasonable to say that the
Local Group contains a massive diffused dark matter component as well \citep{1959ApJ...130..705K}.
Unfortunately, some parameters of the system have not been measured with the adequate accuracy. We
accept the following values in this letter: the  radius of the Solar System orbit
$l_{\odot}=8$~{kpc}, the Milky Way mass $M_{MW}=10^{12} M_\odot$, the Andromeda galaxy mass
$M_{31}=1.6\times 10^{12} M_\odot$, the distance between them $d=750$~{kpc} \citep{collision}. The
tangential components of the velocities of even some massive members of the Local Group are also
not quite explored, and we know almost nothing about the distribution and dynamical parameters of
the diffused component: so the investigation of its motion in the complex gravitational field of
several bodies is quite a difficult and underdefined task. However, our aim is much simpler: we
would like to model just the process of the envelope dark matter penetration towards the Solar
System. Taking into account all the above-mentioned uncertainties, we will try to construct a
simple toy model, that does not claim to describe all properties of the Local Group, allowing us,
however, to estimate the density and the velocity distribution of the extragalactic dark matter
near the Solar System.

Let us consider the following model: the system is stationary and spherically-symmetric. Specific
angular momentum $\mu\equiv[\vec v\times\vec r]$ and maximum radius $r_0$ the particle moves from
the centre remain constant for each particle in such a system, and gravitational potential
$\phi(r)$ depends only on $r$. $r_0$ of the particles belonging to the envelope lie in some
interval $[r_{in},r_{out}]$. We accept $r_{in}=300$~{kpc}, which approximately corresponds to the
size of the Milky Way Roche lobe in system Milky Way - M31, $r_{out}=600$~{kpc}, in accordance with
\citep{collision}. The particles have some distribution $f(r_0)$ over $r_0$ inside
$[r_{in},r_{out}]$; we assume that their specific angular momentum $\mu\equiv[\vec v\times\vec r]$
has Gaussian distribution. So the overall distribution (i.e. the mass $dm$ of particles in some
interval $dr_0 d\mu$) is:
\begin{equation}
 \label{14a1}
dm = f(r_0) \frac{2\mu}{\alpha^2} \exp\left(-\frac{\mu^2}{\alpha^2}\right)\; d\mu dr_0, \qquad
r_0\in [r_{in},r_{out}]
\end{equation}
where $\alpha$ is, generally speaking, a function of $r_0$. We accept the envelope mass
$M_{env}=\int^{r_{out}}_{r_{in}}\! f(r_0) dr_0 = 10^{12} M_\odot$. This value is noticeably smaller
than the total mass of the diffused component, which is estimated as $\sim M_{MW}+ M_{31}=2.6\times
10^{12} M_\odot$ \citep{collision}. We allow for, however, that the main part of this substance
surrounds and accretes on the Andromeda galaxy, and take only $M_{MW}/(M_{MW}+ M_{31})$ of the
total mass, so we assume that the mass should be divided proportionally to the Roche lobe areas of
the components.

At first glance it would seem that the above-stated model is completely unusable to describe the
dark matter motion in the Local Group: the gravitational field of the system by no means can be
considered as central on the scale $\sim 600$~{kpc}, because of huge perturbations from other group
members, especially from the Andromeda galaxy. However, we are only interested in the envelope dark
matter penetration towards the Solar System, and this process is totally defined by the angular
momenta of the particles: only the particles with very small momenta can reach the Earth. The
motion of the particles near the Solar System is almost unaffected by M31 ($\lo\ll d$) and may well
be described by the above-mentioned model. When a particle moves from the envelope to the Earth,
its angular momentum is, of course, strongly influenced by the tidal perturbations. However, we
almost do not know the momentum distribution of the particles in the envelope. Therefore equation
(\ref{14a1}) may be thought of as describing the resultant distribution of the falling particles
with regard to the perturbations from the other members of the Local Group. Moreover, as we will
see, the shape of angular momentum distribution is not very important: we actually use only the
value at $\mu=0$. As for perturbations of the particle energy, they are of the order of
$GM_{31}/d$, i.e. always small. This is a result of the fact that the main part of the particle
acceleration takes place deep in the  Milky Way, where the gravitational field is much stronger.
\looseness=-1

Now we should find the particle distribution inside radius $r_{in}$. A very similar task has been
studied extensively in \citep{2012arXiv1205.4302B}. We cite here only the results adaptable to our
work, skipping the complete derivation. The exact distribution inside $r_{in}$ depends on $r$ and
is equal to:
 \begin{equation}
 \label{14a3}
\rho =\int\limits^{r_{out}}_{r_{in}}\!\!\int^{\mm}_0\!\!\!\!\!\frac{f(r_0) r_0\mu
\exp\left(-\mu^2/\alpha^2\right) d\mu dr_0}{2 \pi r \alpha^2(r_0) T(r_0,\mu) \sqrt{r_0^2-r^2}
\sqrt{\mm^2-\mu^2}}
\end{equation}
Here $T(r_0,\mu)$ is the half-period of a particle with maximal radius $r_0$ and specific angular
momentum $\mu$, i.e. the time it takes for the particle to fall from its maximal radius to the
minimal one, and $\mm$ is the maximum angular momentum of a particle wherewith it can reach radius
$r$
\begin{equation}
 \label{14a4}
 \mm^2 = 2 (\phi(r_0)-\phi(r)) \left(\frac{1}{r^2}-\frac{1}{r_0^2}\right)^{-1}
\end{equation}
Our concern is only with the particle distribution at $r=l_{\odot}$. Since $l_{\odot}\ll
r_{in}<r_0$, we can simplify the above equations
 \begin{equation}
 \label{14a5}
\rho =\int\limits^{r_{out}}_{r_{in}}\!\!\int^{\mm}_0\!\!\!\!\!\frac{f(r_0) \mu
\exp\left(-\mu^2/\alpha^2\right) d\mu dr_0}{2 \pi l_{\odot} \alpha^2(r_0) T(r_0)
\sqrt{\mm^2-\mu^2}}
\end{equation}
\begin{equation}
 \label{14a6}
 \mm(l_{\odot}) = l_{\odot} \sqrt{2 (\phi(r_0)-\phi(l_{\odot}))}
\end{equation}
Here we took into account that for $\mu\in [0,\mm(l_{\odot})]$ period $T(r_0,\mu)$ is almost
independent on $\mu$ ($T(r_0,\mu)\simeq T(r_0,0)\equiv T(r_0)$, see \citep{2012arXiv1205.4302B} for
details).

Equation (\ref{14a5}) can be significantly simplified, if we take into account that $\alpha(r_0)$
is physically constrained. On the one hand, $\alpha(r_{out})$ hardly can be higher, than
\begin{equation}
 \label{14a7}
 \alpha(r_{out})=\frac13 r_{out} \sqrt{\frac{2G(M_{MW}+M_{env})}{r_{out}}}
\end{equation}
since in the opposite case a significant fraction of the particles would have the speed above the
escape velocity at this radius. On the other hand, numerical simulations \citep{mo09} show that the
root-mean-square angular momentum of the particles should be quite high and close to upper limit
(\ref{14a7}). Though there is rather strong evidence that $\alpha$ of the particles of our Galaxy
is much lower \citep{2011MNRAS.417L..83B}, we will use value (\ref{14a7}) in our calculations,
since the density of the extragalactic dark matter grows with decreasing of $\alpha$, and we take
the maximum possible value in order to obtain a conservative estimate. The dependence of $\alpha$
on $r_0$ is not well known, and we will presume it to be power-law
\begin{equation}
 \label{14a8}
 \alpha(r_{0})=\alpha(r_{out})\left(\frac{r_0}{r_{out}}\right)^i
\end{equation}
However, we only need much softer condition $\alpha(r_0)\ge\mm(\lo)$ to simplify (\ref{14a5}). It
means that the tangential velocity dispersion in the envelope is supposed to be higher than
$v_{esc}\frac{\lo}{r_{in}}\simeq 16$~{km/s}. Such an assumption seems quite natural. If
$\alpha(r_0)\ge\mm(\lo)$, we can simplify (\ref{14a5}) as

 \begin{equation}
 \label{14a9}
\rho =\int\limits^{r_{out}}_{r_{in}}\!\!\int^{\mm}_0\!\!\!\!\!\frac{f(r_0) \mu
 d\mu dr_0}{2 \pi l_{\odot} \alpha^2(r_0) T(r_0)
\sqrt{\mm^2-\mu^2}}
\end{equation}

 Now we should ascertain the velocity distribution of the particles.
Let us denote the tangential and radial components, and the total velocity of a particle at
$r=l_\odot$ by $u_\tau$, $u_r$, and $u$ respectively. We can use $u_\tau$ and $u$ instead of $r_0$,
$\mu$. Indeed, $\mu=u_\tau l_\odot$, $u=\sqrt{2(\phi(r_0)-\phi(\lo))}$, $\mm=u l_\odot$. Let us
consider the particles with the same $u$ (i.e., with the same $r_0$)  and find their angular
distribution. An element $d\Omega$ of the solid angle in the phase space is equal to
\begin{equation}
 \label{14a10}
 d\Omega=\frac{u}{u_r} \frac{2\pi u_\tau du_\tau}{4\pi u^2}=\frac{\mu d\mu}{2\lo \sqrt{\mm^2-\mu^2} \sqrt{2(\phi(r_0)-\phi(\lo))}}
\end{equation}
Substituting this equation to (\ref{14a9}), we obtain
 \begin{equation}
 \label{14a11}
\rho =\int\limits^{r_{out}}_{r_{in}}\!\!\frac{f(r_0) \sqrt{2(\phi(r_0)-\phi(\lo))} dr_0}{\pi
\alpha^2(r_0) T(r_0)} \int\!\! d\Omega
\end{equation}
As we can see, the particle distribution depends only on $r_0$ (i.e., on velocity magnitude $u$),
and is independent on the direction. So the distribution is isotropic. $u$ and $r_0$ are bound by a
one-to-one relation
 \begin{equation}
 \label{14a12}
u=\sqrt{2(\phi(r_0)-\phi(\lo))}\qquad
du=\frac{\left(\frac{d\phi(r_0)}{dr_0}\right)dr_0}{\sqrt{2(\phi(r_0)-\phi(\lo))}}
\end{equation}
We can substitute this equation to (\ref{14a11}) and take into account that $\int\! d\Omega=4\pi$.
 \begin{equation}
 \label{14a13}
\rho =\int\limits^{\sqrt{2(\phi(r_{out})-\phi(\lo))}}_{\sqrt{2(\phi(r_{in})-\phi(\lo))}}\! \frac{8
f(r_0) (\phi(r_0)-\phi(\lo))}{ \alpha^2(r_0) T(r_0) (d\phi(r_0)/dr_0)} du
\end{equation}
Hence the velocity distribution of the dark matter from the envelope is isotropic near the Solar
System; equations (\ref{14a12}) and (\ref{14a13}) totally define its density and momentum
distribution.

\begin{figure}
 \resizebox{\hsize}{!}{\includegraphics[angle=0]{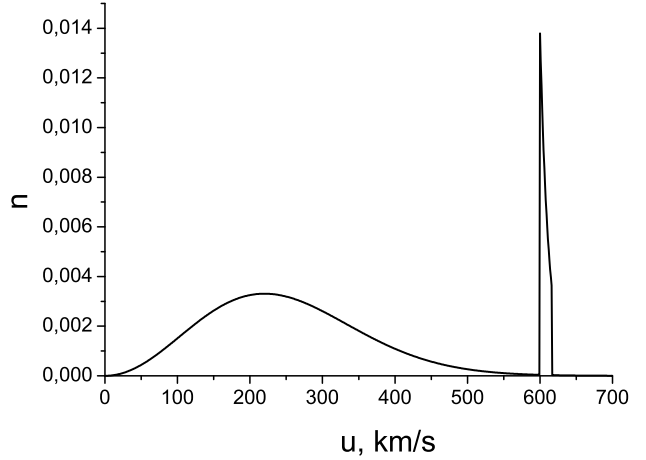}}
\caption{The normalized velocity distribution of dark matter particles near the Earth in the frame
of reference that does not rotate around the Galaxy centre. The distribution of the galactic DMPs
is supposed to be Maxwell (\ref{14a23}). The extragalactic component gives a narrow high peak near
$600$~{km/s}.}
 \label{fig1}
\end{figure}

In order to complete the solution, we should define functions $\phi(r_0)$, $T(r_0)$, and $f(r_0)$.
We will assume $f(r_0)$ to be a power-law function with some index $j$. Since $\int\! f(r_0) dr_0=
M_{env}$,
 \begin{equation}
 \label{14a14}
f(r_0)=\frac{(j+1)M_{env}}{r_{out}-r_{in}}\left(\frac{r_0}{r_{out}-r_{in}}\right)^{j}
\end{equation}
We accept the mass distribution at $r>r_{in}$ to be
 \begin{equation}
 \label{14a15}
M(r)\simeq M_{MW}+M_{env}\left(\frac{r-r_{in}}{r_{out}-r_{in}}\right)^{j+1}
\end{equation}
This equation is not quite true, because in fact the particles contribute as well to the mass
inside $r_{in}$. However, our calculation is estimative, and we may disregard rather a small
deviations from (\ref{14a15}). With the same accuracy $T(r_0)$ may be thought of as a power-law
function. One can readily see that $T(r_0)\propto r^{1-\frac{j}{2}}$ if $M(r)\propto r^{j+1}$.
Though the mass distribution is not quite power-law in our case, we presume
 \begin{equation}
 \label{14a16}
T(r_0)=T(r_{in})\left(\frac{r_0}{r_{in}}\right)^{1-\frac{j}{2}}
\end{equation}
and $T(r_{in})=10^{17}~s$, which is slightly more, than the time necessary for a body to fall on
point mass $10^{12} M_\odot$ from $300$~{kpc} with no initial velocity. Function
$\frac{d\phi(r_0)}{dr_0}=\frac{G M(r_0)}{r^2_0}$ is totally defined by (\ref{14a12}).
 \begin{equation}
 \label{14a17}
\frac{d\phi(r_0)}{dr_0}=\frac{G}{r^2_0}\left[M_{MW}+M_{env}\left(\frac{r_0-r_{in}}{r_{out}-r_{in}}\right)^{j+1}\right]
\end{equation}
As for function $\sqrt{2(\phi(r_0)-\phi(\lo))}$, it remains almost constant for
$r_0\in[r_{in};r_{out}]$ owing to the smallness of $|\phi(r_{out})-\phi(r_{in})|$ as compared with
$|\phi(r_{in})-\phi(r_{\lo})|$. Therefore we can approximate $\sqrt{2(\phi(r_0)-\phi(\lo))}\simeq
\sqrt{2(\phi(r_{in})-\phi(\lo))}\equiv V$. Hereafter we accept $V=600$~{km/s}. The velocities of
all the particles lie in a very narrow interval $\Delta V$
 \begin{equation}
 \label{14a18}
\Delta V=\sqrt{V^2+2(\phi(r_{out})-\phi(r_{in}))} - V
\end{equation}
Now we should substitute equations (\ref{14a8}), (\ref{14a16}), (\ref{14a14}) for $\alpha(r_0)$,
$T(r_0)$, $f(r_0)$ respectively and (\ref{14a7}) for $\alpha(r_{out})$ into equation (\ref{14a13}).
It is convenient to introduce $k\equiv r_{out}/r_{in} = 2$. After some trivial calculations we
obtain
 \begin{equation}
 \label{14a19}
\rho=9\frac{(j+1)}{(\frac32j - 2i)}\frac{k^{2j}(k^{\frac32j-2i}-1)}{(k-1)^{j+1}}\dfrac{V}{G
T(r_{in}) r_{out}}
\end{equation}
The shape of velocity distribution function $n$ in the phase space is given by
 \begin{equation}
 \label{14a20}
n(u)\propto \frac{du^3}{4\pi u^2} r^{\frac32j-2i-1}_0 \left(\frac{dr_0}{du}\right)
\end{equation}
where we should substitute $r_0$ by $u\in[V,V+\Delta V]$ with the help of equations (\ref{14a14})
and (\ref{14a17}). Equations (\ref{14a19}), (\ref{14a20}) completely determine the solution of the
task.

\section{Results and discussion}
Not much is known about the dark matter distribution in the envelope. On the other hand, as we can
see from equation (\ref{14a19}), the result is not strongly dependent on the choice of $i$ and $j$
(as a consequence of the relative smallness of ratio $r_{out}/r_{in}$). It seems reasonable to
choose $i$ and $j$ by analogy with the well-known isothermal halo solution ($dM/dr={\it const}$ and
the Maxwell DMP velocity distribution with a temperature, constant over the halo), which
corresponds to $i=1$, $j=0$. Substituting these values in combination with $T(r_{in})$,
$\alpha(r_{in})$, and $V$ to (\ref{14a19}), we obtain the density of the extragalactic dark matter
near the Earth $\rho= 3.7\times 10^{-2}$~{GeV/cm$^3$}. The speed distribution is notably narrow:
the absolute values of all particles fall within $\Delta V\simeq 16$~{km/s}. As we have already
mentioned, this feature is a consequence of the smallness of $|\phi(r_{out})-\phi(r_{in})|$ as
compared with $-\phi(\lo)$. Therefore, two properties of the velocity distribution are
model-independent: the speeds of extragalactic DMPs from the envelope lie in a narrow range, and
their angular distribution is isotropic.

The density of the extragalactic dark matter turns out to be fairly high: $3.7\times
10^{-2}$~{GeV/cm$^3$} is more than $12$\% of the total dark matter density near the Earth $\simeq
0.3$~{GeV/cm$^3$} \citep{gorbrub1}. This brings up a question: How reliable is the estimation?
Above we have already discussed the approximation of the system by a spherically symmetric model
and found it acceptable. The premise that $f(r_0)$ terminates abruptly at $r_{in}$ and $r_{out}$ is
also unphysical. Undoubtedly, our result is assessed; however, it cannot be called optimistic.
Indeed, as the dimensional method shows, for any envelope model the density of the extragalactic
dark matter is, with an accuracy of a numerical factor, equal to
 \begin{equation}
 \label{14a21}
\rho\propto \frac{M_{env} v_{esc}}{\langle\alpha\rangle^2 \langle T\rangle}
\end{equation}
where $\langle\alpha\rangle$ and $\langle T\rangle$ are the average values of the respective
quantities. Our calculations confirm this dependence: it can be easily obtained from (\ref{14a13}).
$v_{esc}$ is almost independent on the model choice. $\langle T\rangle$ is essentially defined by
the size of the Milky Way Roche lobe, and thus is also more or less model-independent. The main
source of the uncertainty is envelope mass $M_{env}$. We proceed from the assumption of
\citet{collision} that $M_{env}$ is approximately equal to the masses of the galaxies of the Local
Group. We used the highest possible value (\ref{14a7}) for $\alpha$: if $\alpha$ was higher, a
significant part of the envelope would rapidly evaporate. Since $\rho\propto\alpha^{-2}$, this
choice is conservative.

Thus there are two possible situations, when (\ref{14a19}) significantly overestimates the density
of the extragalactic dark matter. It may be so, if the envelope mass is in fact much lower than the
masses of the Local Group member galaxies. The strong overestimation may also appear, if the
angular momentum distribution of the envelope DMPs differs greatly from the Gaussian (\ref{14a1}),
i.e., almost all the particles have circular orbits. Such a supposition seems highly improbable.
First of all, it is in sharp contrast to N-body simulation results \citep{mo09}. There is also a
good indirect counterargument: the largest Local Group member M31 has quite low angular momentum
and, consequently, very oblong orbit \citep{1959ApJ...130..705K}. It is plausible that Milky Way
and M31 will finally experience a central collision. Thus the presence in the diffuse component of
the Local Group of a bulk of dark matter particles that have very oblong orbits and can reach the
Earth seems quite possible.

\begin{figure}
 \resizebox{\hsize}{!}{\includegraphics[angle=0]{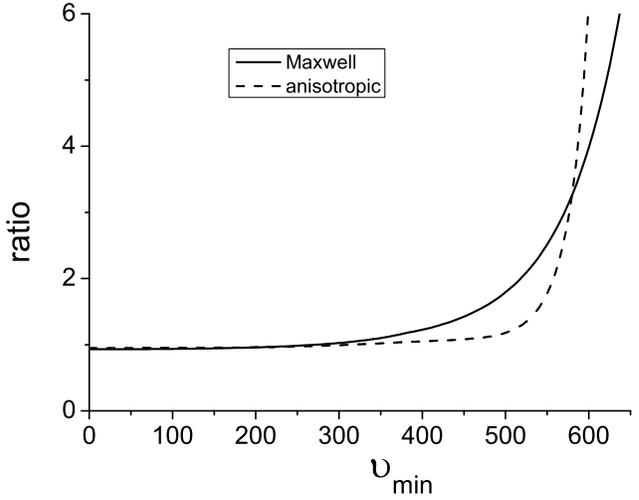}}
\caption{The ratio of the direct detection signal produced by the mixture $\sim 12.3\%$ of
extragalactic component and $87.7\%$ of the galactic DMPs to the signal produced by pure galactic
dark matter. The fraction and velocity distribution of the extragalactic component were calculated
in accordance with (\ref{14a19}). We considered two models of the velocity distribution of the
galactic DMPs: Maxwell (\ref{14a23}) (solid line) and anisotropic (\ref{14a24}) (dashed line). The
extragalactic dark matter almost does not affect the signal, if $\upsilon_{min}<300$~{km/s}, but
totally dominates above $450-500$~{km/s}.}
 \label{fig2}
\end{figure}

 The $12\%$ extragalactic component of the total dark matter density can be especially
important for the direct dark matter search. The direct search is based on the detection of the
collisions of dark matter particles with nuclei of the target. The signal is sensitive to the
velocity distribution: roughly speaking \cite{Belanger}, it is proportional to
\begin{equation}
I(\upsilon_{min})=\int^\infty_{\upsilon_{min}}\frac{\tilde n(\upsilon)}{\upsilon}\; d^3
\vec\upsilon
 \label{11b8}
 \end{equation}
Here  $\tilde n(\upsilon)$ is the distribution in the Earth's frame of reference: it should be
obtained from (\ref{14a20}), (\ref{14a23}), or (\ref{14a24}) by a Galilean transformation.
$\upsilon_{min}$ is the minimal DMP speed, to which the detector is sensitive (see details in
\citet{Belanger}).
 \begin{equation}
 \label{14a22}
\upsilon^2_{min}\simeq \frac{E_A}{2} \frac{(m_\chi+m_A)^2}{m_A m^2_\chi}
\end{equation}
where $m_\chi$ and $m_A$ are the DMP and the detector nucleus mass respectively, $E_A$ is the
detector activation energy, depending on its construction. In order to estimate the influence of
the extragalactic dark matter to the direct detection signal, we should define a model for the
velocity distribution of the galactic component. The Maxwell distribution is now routinely used,
mainly because of its simplicity:
\begin{equation}
n(u) = \frac{1}{(\sqrt{\pi} \upsilon_{\odot})^3} \exp \left(- \frac{u^2}{\upsilon_{\odot}^2}\right)
 \label{14a23}
\end{equation}
$\upsilon_{\odot}$ is the orbital speed of the Solar System. There are strong reasons to suppose,
however, that the distribution of the galactic DMPs is strongly anisotropic and looks like
\begin{equation}
n(u)=
\frac{\exp\left(-\dfrac{u^2_\tau}{2\sigma^2_0}\right)}{2\pi^2\sigma^2_0\sqrt{u^2_{max}-u_r^2}}
 \label{14a24}
 \end{equation}
where $u_r\in [-\upsilon_{max};\upsilon_{max}]$, $u_{max}\simeq 560$~{km/s}, $\sigma_0=80$~{km/s},
$u_r$ and $u_\tau$ are radial and tangential components of the particle velocity, respectively
\citep{2011MNRAS.417L..83B}.

 We calculated the signal produced by mixture of $3.7\times 10^{-2}$~{GeV/cm$^3$} of the
extragalactic dark matter and $0.263$~{GeV/cm$^3$} of the galactic one and divided it by the signal
produced by the pure galactic dark matter with the same total density ($0.3$~{GeV/cm$^3$}). We used
both models of the galactic DMP distribution: Maxwell (\ref{14a23}) and anisotropic (\ref{14a24}).
The two ratios are represented in Fig.~\ref{fig2} by the solid line (Maxwell velocity distribution
of the galactic dark matter particles) and by the dashed line (anisotropic velocity distribution).
One can see that the signal is scarcely affected by the presence of the extragalactic component, if
$\upsilon_{max}<300$~{km/s}. However, the situation drastically changes for higher
$\upsilon_{max}$; if $\upsilon_{max}$ is larger, than $450-500$~{km/s}, the extragalactic signal
dominates. This is not particularly surprising: all the extragalactic particles are faster than
$600$~{km/s}, while the number of the galactic particles rapidly drops above $\sim 450$~{km/s}.
Hence the impact of the extragalactic dark matter to the direct detection signal can be very
important, especially if the DMP mass $m_{\chi}$ is small. Indeed, if $m_{\chi}$ is small,
$\upsilon_{max}$ is high (\ref{14a22}), i.e., we can detect only the fastest DMPs. For instance,
DAMA collaboration reports \citep{dama} about the detection of a signal produced by $\sim 10$~GeV
weakly interacting massive particles. We shall not discuss here the question of the nature of the
signal (other detectors do not confirm the result \citep{xenon}). It should be recorded, however,
that $\upsilon_{max}>450$~{km/s} for the majority of detectors, if the DMP is so light, i.e., the
extragalactic component signal should totally dominate.

Notice that we should, strictly speaking, have cut distributions (\ref{14a23}) and (\ref{14a24}) at
$u=v_{esc}$. However, the fraction of the particles with $u>v_{esc}$ is negligible for both the
distributions, and the cutting would hardly affect the result; the impact of the extragalactic
component would be even slightly higher.

 In conclusion, let us briefly consider the extragalactic dark matter that does not belong to the
Local Group. As recent astronomical observations imply \citep{karach11}, the dark matter of the
Virgo Supercluster, in addition to galaxies and their groups, forms a large diffuse component. We
do not know its velocity and space distributions, but it seems reasonable to assume that the dark
matter is distributed more or less uniformly, and the velocity dispersion of the DMPs is comparable
with that of the observable members of the Supercluster ($v_{\infty}\sim 500$~{km/s}). The
measurements estimate the average density of the diffuse component as $\rho\sim
10^{-6}$~{GeV/cm$^3$}. The gravitational field of the Local Group should increase this quantity
near the Earth. However, we can roughly estimate the enhancement as $1+v^2_{esc}/v^2_{\infty}$,
where $v_{esc}\simeq 650$~{km/s} is the escape velocity from the Solar System orbit
\citep{2012MNRAS.420..590B}. Thus the density of the Supercluster dark matter is approximately $3$
times higher near the Solar System, but yet hardly exceeds $10^{-5}$~{GeV/cm$^3$}. This value is so
low, that it may scarcely be of interest for modern experiments. On the other hand, the
Supercluster DMPs are particularly energetic ($v>1000$~{km/s}) and hence may give a very
characteristic signal.

To summarize:

1) The particles of the diffuse component of the Local Group are apt to contribute $\gtrsim 10\%$
to the total dark matter density near the Earth.

2) The particle speeds are $\sim 600$~{km/s}, i.e. they are much faster than the galactic DMPs. The
particles have isotropic velocity distribution (perhaps, in contrast to the galactic dark
 matter); their speed distribution is very narrow ($\Delta V\sim 20$~{km/s}).

3) The extragalactic dark matter should give a significant contribution to the direct detection
 signal. If the detector is sensitive only to the fast particles ($v> 450$~{km/s}), the
 signal may even dominate.

4) The density of other types of the extragalactic dark matter (for instance, of the DMPs forming
the diffuse component of the Virgo Supercluster) should be relatively small and comparable with the
average dark matter density of the Universe. However, these particles can generate anomaly
high-energy collisions in direct dark matter detectors.

Financial support by Bundesministerium f\"ur Bildung und Forschung through DESY-PT, grant 05A11IPA,
is gratefully acknowledged. BMBF assumes no responsibility for the contents of this publication.

\label{lastpage}
\end{document}